\documentclass[useAMS,usenatbib]{mn2e}
\usepackage{natbibmnfix}
\usepackage{graphicx}
\usepackage{subfigure}
\usepackage{mathrsfs,amssymb}
\usepackage{amsmath}
\usepackage{natbib}
\usepackage{color}
\usepackage{ulem}
\usepackage{url}

\newcommand{\aj}{\rm AJ}                   
\newcommand{\apj}{\rm {ApJ}}                 
\newcommand{\apjl}{\rm {ApJ}}                

\newcommand{\mnras}{\rm {MNRAS}}             
\def\gtsima{$\; \buildrel > \over \sim \;$}
\def\ltsima{$\; \buildrel < \over \sim \;$}
\def\gsim{\lower.5ex\hbox{\gtsima}}
\def\lsim{\lower.5ex\hbox{\ltsima}}
\def\simgt{\lower.5ex\hbox{\gtsima}}
\def\simlt{\lower.5ex\hbox{\ltsima}}
\def\simpr{\lower.5ex\hbox{\prosima}}

\begin{document}

\title
[Ultra-faint galaxies]
{Ultra-faint high-redshift galaxies in the Frontier Fields}
\author[Yue et al.]{B. Yue$^{1}$, A. Ferrara$^{1,2}$, E. Vanzella$^{3}$, R.
  Salvaterra$^{4}$\\
$^1$Scuola Normale Superiore, Piazza dei Cavalieri 7, I-56126 Pisa, Italy\\
$^2$Kavli IPMU (WPI), Todai Institutes for Advanced Study, the University of Tokyo, Kashiwa, Chiba 277-8583, Japan\\	
$^3$INAF, Osservatorio Astronomico di Bologna, Via Ranzani 1, I-40127, Bologna, Italy\\
$^4$INAF, IASF Milano, via E. Bassini 15, I-20133 Milano, Italy \\
}
\maketitle

\begin{abstract}
By combining cosmological simulations with Frontier Fields project lens models we find that, in the most optimistic case, 
galaxies as faint as $m \approx 33 - 34$  (AB magnitude at $1.6~\rm \mu m$) can be detected in the Frontier Fields.
Such faint galaxies are hosted by dark matter halos of mass $\sim10^9~M_\odot$ and dominate the ionizing photon budget 
over currently observed bright galaxies, thus allowing for the first time the investigation of the dominant reionization sources. 
In addition, the observed number of these galaxies can be used to constrain the role of feedback in suppressing star formation 
in small halos: for example, if galaxy formation is suppressed in halos with circular velocity $v_c < 50$~km s$^{-1}$, galaxies fainter than $m=31$ should not be detected in the FFs.
\end{abstract}


\section{introduction}

Observing high-$z$ faint galaxies challenges the capability of current telescopes. Among high-$z$ objects, usually the brightest galaxies that formed in rare, massive halos are more easily detected. Nowadays, observations of the high-$z$ galaxy populations 
have reached $\sim 0.1 L^\star$ \citep{2007ApJ...670..928B,2010ApJ...709L..16O, 2010MNRAS.403..960M,2010ApJ...725.1587B,2011ApJ...737...90B, 2012ApJ...745..110O,2013ApJ...768..196S,2013ApJ...773...75O,
2013arXiv1309.2280O}. However, it is believed that the majority of high-$z$ galaxies are dwarf systems which are too faint to be detected by current telescopes \citep{2011MNRAS.414..847S,2013MNRAS.434.1486D}. Accessing such a faint population of early galaxies is a compelling task as known high-$z$ galaxies cannot provide enough ionizing photons to complete the reionization process and match the measured Thomson optical depth. It is commonly believed that  faint galaxies are required in order to provide the missing ionizing power \citep{2007MNRAS.380L...6C,2011MNRAS.414.1455L, 2012MNRAS.420.1606J, 2012ApJ...752L...5B,2012ApJ...758...93F}.

An exciting possibility to discover and study these faint galaxies is represented by gravitational lensing by relatively nearby clusters, that could magnify the brightness of objects behind the 
lens \citep{2009ApJ...690.1764B,2012MNRAS.423.2829O,
2012MNRAS.427.2212Z,2013ApJ...762...32C, Coe14,2014ApJ...783L..12V}. 
A major effort along these lines is currently undertaken by the HST Frontier Field (FF) project\footnote{\url{http://www.stsci.edu/hst/campaigns/frontier-fields/}}. This project aims to
perform a photometric survey of areas behind six clusters and six flanking blank fields, achieving a detection threshold deeper than any previous 
such kind surveys: $H=28.7$ magnitude (5$\sigma$ detection for point sources)
in the $H$ band\footnote{\url{http://www.stsci.edu/hst/campaigns/frontier-fields/HST-Survey}}, implying that  objects fainter than $m=28.7+2.5 \log \mu$,
where $\mu$ is the magnification, will be detected.
\footnote{Throughout this {\it Letter}, we 
use AB magnitudes;
``$m$" denotes the intrinsic apparent magnitude, i.e. before lensing;
``$H$" denotes the observed apparent magnitude, including the lensing effect.}
The primal observations for the cluster Abell 2744 have already been analyzed \citep{2013arXiv1311.7670A,2014arXiv1401.8263L}.

In this {\it Letter}, we aim to characterize the properties of the faint galaxy populations detectable by the FFs project, based on the numerical simulations of \citet{2011MNRAS.414..847S}. We also investigate the possibility to use the number counts of the lensed faint galaxies to constrain the suppression of the galaxy formation in small 
dark matter 
halos by feedback processes.

\section{Faint galaxy number counts}\label{number}

The number of galaxies with intrinsic apparent magnitude larger than $m$ between redshift $z_1$ and $z_2$ observed after lensing is \citep{2000ISASS..14..171Y}
\begin{equation}
N_{\rm g}(> m)=\int_{z_1}^{z_2}dz'r^2\frac{dr}{dz'}
\iint\limits_\Omega
\frac{d\theta_yd\theta_x}{\mu(\theta_x,\theta_y,z')}\int_{L(m_{\rm lim})}^{L(m)}\Phi(L',z')dL',
\label{Ng}
\end{equation}
where the integration covers the whole field of view (FOV) $\Omega$;
$L(m)$ is the luminosity corresponds to apparent magnitude $m$,
and $\Phi$ is the luminosity function (LF) of galaxies.
$m_{\rm lim} = H_{\rm lim}+2.5\log\mu(\theta_x,\theta_y,z')$, 
while $H_{\rm lim}$ is the detection limit of the experiment. 
The measured number density is the combination of the surface number density dilution ($\mu$ in the denominator)
and the brightness boost ($\mu$ in $m_{\rm lim}$).

We use the LFs obtained from the \citet{2011MNRAS.414..847S} SPH simulations. 
The simulated volume has a linear (comoving) size
$L = 10 h^{-1}$ Mpc with $N_p = 2\times256^3$ (dark+baryonic) particles, corresponding to a dark matter (baryonic) particle
mass of $M_p = 3.62 \times 10^6 h^{-1} M_\odot~(6.83 \times 10^5 h^{-1} M_\odot)$; the corresponding force resolution is 2 kpc.
We refer the reader to the original paper for additional simulation details. 
The simulated LFs match very well the observed ones between $z \sim 5 - 10$. In the following, we will always assume that the LF is observed at a rest-frame wavelength $1.6/(1+z)~\rm \mu m$, while the apparent magnitude is measured in the H-band  ($1.6~\rm \mu m$).
Therefore $H_{\rm lim} = 28.7$.
From the simulation outputs we build a $\Phi(L',z')$ grid and interpolate where necessary.
To correct for numerical resolution effects plaguing very faint ($M_{\rm AB}\simgt-15$) galaxies hosted by halos with 
$\sim 150$ dark matter particles, we use a normalized power-law LF with slope 
-2 \citep{2013MNRAS.429.2718S} at this faint part.
Moreover, to account for radiative feedback effects suppressing star formation in $T_{\rm vir} \le 10^4$~K \citep{Xu13}, we 
cut the power-law LF at a magnitude $M_{\rm AB}^{\rm cut}\sim-11$ to $-13.5$ between $z = 10 - 5$ at which the number of galaxies brighter than $M_{\rm AB}^{\rm cut}$  equals the number of halos with $T_{\rm vir} \ge 10^4$~K, taken from 
\citet{2001MNRAS.323....1S}.

\begin{figure}
\centering{
\includegraphics[scale=0.4]{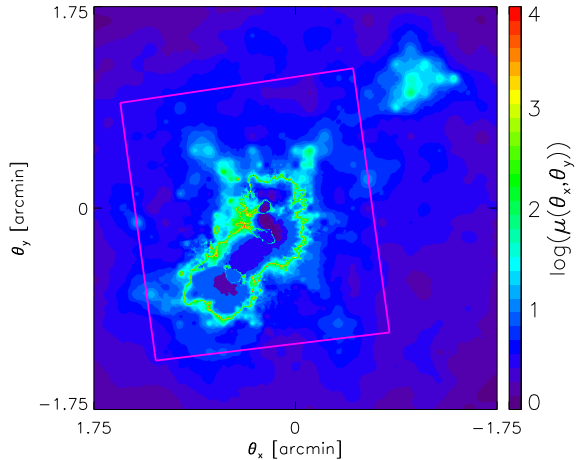}
\caption{Magnification map for the cluster Abell 2744 with sources located at $z=5$.
The HST/WFC3 FOV is marked by magenta lines.}
\label{magnifmap}
}
\end{figure}

To compute the magnification, we use the lens models released on the website of the FFs project\footnote{\url{http://archive.stsci.edu/prepds/frontier/lensmodels/}}. For each lensing cluster, six lens models are provided; the fiducial results are based on the Bradac model. However, we have checked that other models do not qualitatively change our conclusions. As an example, we show the magnification map for Abell 2744 with sources located at $z=5$ in Fig. \ref{magnifmap}. The FOV of HST/WFC3 is marked by magenta lines.

\begin{figure}
\centering{
\subfigure{\includegraphics[scale=0.4]{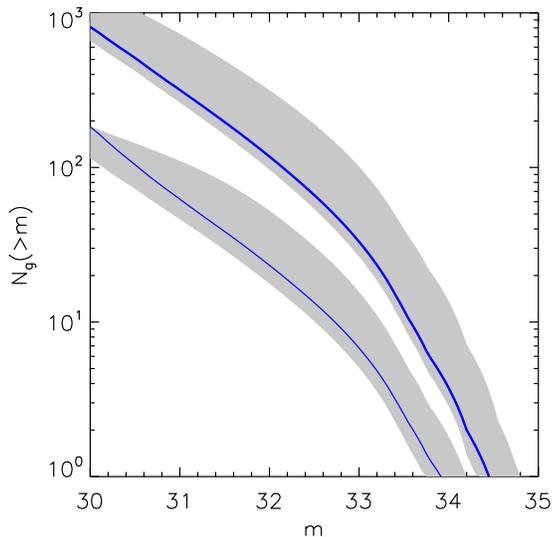}}
\subfigure{\includegraphics[scale=0.4]{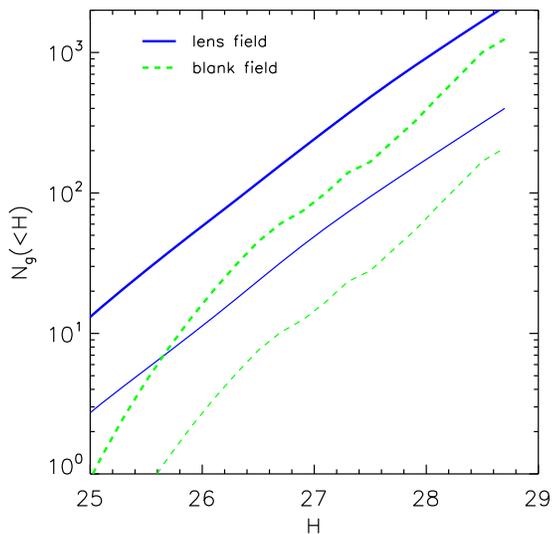}}
\caption{
{\it Upper panel:} Number of galaxies fainter than \textit{intrinsic} magnitude $m$ detectable by 
the FFs project and variance due to lens models.
{\it Bottom:} Number of galaxies brighter than \textit{measured} magnitude $H$ in the cluster (solid) and in the blank (dashed) fields. Thin lines refer to Abell 2744; thick lines include all six lenses.}
\label{N_observed}
}
\end{figure}

We plot the number of observed galaxies {\it fainter} than H-band intrinsic magnitude $m$ 
and have $5 \le z \le 10$ in the upper panel Fig. \ref{N_observed}.
The curves give the results based on the Bradac model, the difference between
lens models is shown by filled regions. Our discussions are based on 
Bradac model hereafter.
In the FFs project, it is possible to detect galaxies as faint as $m \approx 33 - 34$ ($N_g \simgt 1$, i.e. at least one galaxy) behind the 
Abell 2744 (thin lines), or one magnitude deeper if all the six lens clusters are considered (thick lines). 
About 15 such galaxies with $m = 32-33$ between $z=5$ and $z=10$ are expected behind Abell 2744,
while if the survey of all the six lenses is completed, such number increases to  $\sim80$. 
Of course this is the most optimistic case as galaxies are not point sources; considering the incompleteness correction, the actual detection depth for galaxies could be more than one magnitude shallower \citep{2001AJ....122.2190V}.
In the bottom panel of Fig. \ref{N_observed}, we plot the number of observed galaxies in $5 \le z \le 10$, and {\it brighter} than 
H-band {\it observed} magnitude in the lens fields (solid) and in the blank fields (dashed).

The Abell 2744 WFC3 field area ($\sim2.3'\times2.5'$) contains $ \sim 1.5\times10^5$ halos 
above $1.4\times10^8~M_\odot$ in $5 < z < 10$. Among them, about 200 could be observed as 
galaxies with $m \ge 30$; therefore the completeness is $\sim 0.1\%$. Using the Cosmic Variance Calculator
\citep{2008ApJ...676..767T}\footnote{http://casa.colorado.edu/\~trenti/CosmicVariance.html}
we get a relative variance $\sim10\%$. About 15 could be observed as galaxies with 
$m  \ge 32$, yielding a completeness of $\sim0.01\%$ and a relative variance of $\sim30\%$.

\begin{figure}
\centering{
\includegraphics[scale=0.4]{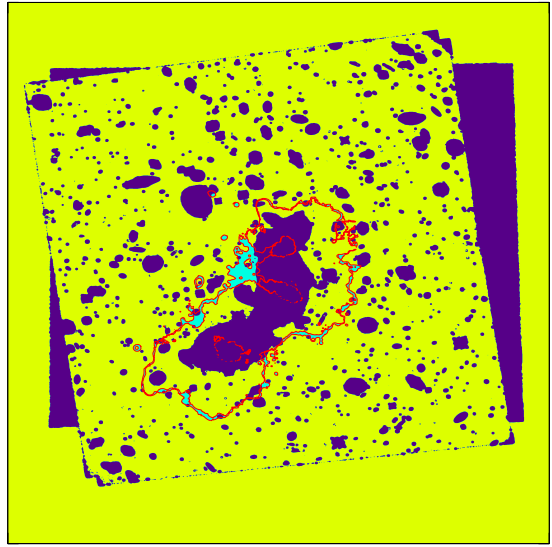}
\caption{The purple areas are regions in the Abell 2744 map where point sources as faint as 
$H=28.7$ (lensed apparent magnitude) are non-detectable. We also show the positions where $\mu > 100$ in the Bradac lens model (narrow regions between the two solid lines) for a source located at $z=5$. 
Cyan regions correspond to areas in which the magnification 
is greater than 100 and source/galaxy as faint as $H=28.7$ can be detected.}
\label{mask}
}
\end{figure}

In practice, however, foreground light contamination from the lensing cluster itself 
may reduce
the detectability. We estimate this effect as follows. We downloaded the latest H-band image of 
Abell 2744 from the FFs website. This image is used as the foreground map (we neglect the fact that some light 
in this map is from the high-$z$ galaxies we are searching for, as these only fill a very small area). The FWHM 
of the PSF is about 0.2$''$; as a result,  for a point source and a Gaussian PSF, the central pixel 
contains about 30\% of the total flux (the size of each pixel is 0.06$''$). Centered on this pixel, we choose a 
region within a certain radius, $R$, and compute the mean, $\langle F \rangle$, and rms, $\sigma$, flux of all 
pixels in this region. This point source is defined as detectable in this search radius when its central flux peak 
is: a) $>5 \sigma$, and b) $>\langle F \rangle$ within this radius.  We use different values of $R$ in the range 
0.1$''$ to 0.6$''$; A point source is defined as detectable only if it is detectable for all values of $R$ in this range. 
Assuming a point source with $H=28.7$, we mask the map areas in which such source is not detectable (purple 
regions in Fig. \ref{mask}). We find that 80\% of the regions with $\mu > 100$ are not masked. For brighter sources, 
this fraction would be even larger. Therefore we conclude that the contamination does not constitute a serious problem 
for the search of ultra-faint galaxies.

\section{Faint galaxy properties}
What are the physical properties of the high-$z$ faint ($m \lesssim 33$) galaxies that will be possibly detected in the FFs?
We are particularly interested in galaxies fainter than $m=30$ and therefore are non-detectable in current deepest 
surveys \citep{2013ApJ...773...75O} without lensing.
The ultra-faint galaxies are grouped in three magnitude bins: $m=30 - 31$, $m=31 - 32$ and $m=32 - 33$,
respectively. Their simulated properties are discussed in the following.

\begin{figure}
\centering{
\includegraphics[scale=0.4]{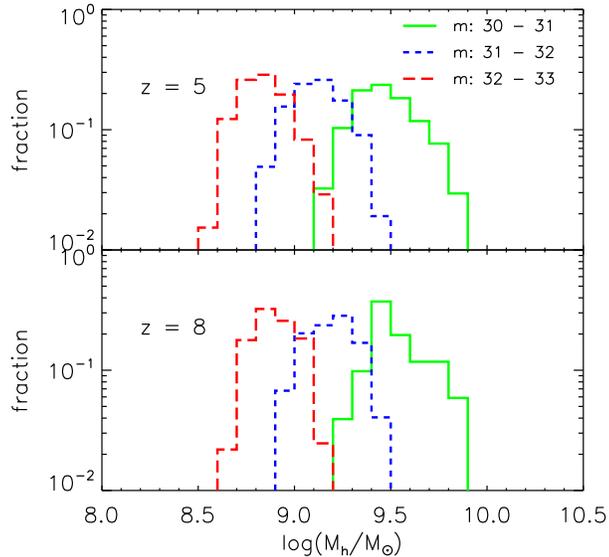}
\caption{The mass distribution of the host halos in various intrinsic apparent magnitude
bins at redshift $z=5$ (up) and $z=8$ (bottom). 
}
\label{halomass}
}
\end{figure}

The mass distribution of the host halos of these galaxies in various magnitude bins is plotted in Fig. \ref{halomass}. The typical halo 
mass depends on brightness, with the brighter galaxies located in larger halos. However, even for galaxies in the $30 - 31$ magnitude bin, the host halo distribution peaks around masses as small as $\approx 10^{9.5}~M_\odot$. At $z=5$,  about 10\% of the galaxies in the $32 - 33$ magnitude bin resides in even smaller halos, $\approx 10^{8.5}~M_\odot$; the corresponding virial temperature of such halos is $T _{\rm vir}\approx 1.5\times 10^4$ K (assuming ionized gas), i.e. very close to the lower boundary ($10^4$ K) of atomic-cooling halos. $T _{\rm vir} \simgt 10^4$ K halos should be strongly affected by photoionization feedback which evaporates their gas and 
strongly suppresses their star formation. Hence, determining the abundance of such halos allows a precise assessment of the 
importance of feedback in the early Universe. We will come back to these implications later on. 

\begin{figure}
\centering{
\includegraphics[scale=0.4]{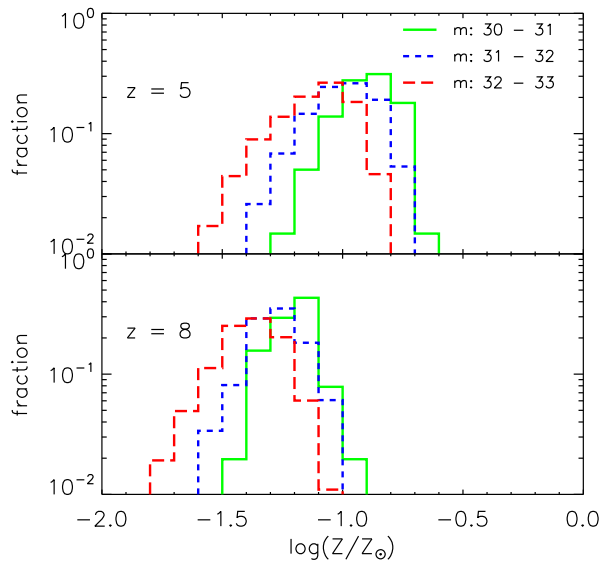}
\caption{Same as Fig. \ref{halomass} but for the stellar metallicity distribution. 
}
\label{metallicity}
}
\end{figure}

Quite interestingly, even these small galaxies are already relatively metal enriched, as can be seen by inspecting Fig. \ref{metallicity}.
At a given redshift, the difference in metallicity trend with magnitude is not as obvious as for the host halo mass.
At $z=5$, the mean ultra-faint galaxy metallicity is $\sim0.1~Z_\odot$, while at $z\sim8$ it is $\sim0.05~Z_\odot$. 
We also find that even in such faint galaxies, there is little 
hope to detect the Pop III star component, as its contribution 
to the H-band luminosity
is limited to only $\approx $1\% of the host galaxies. 
We have also computed the UV spectral slope, $\beta$ 
($L_\lambda \propto \lambda^\beta $),  of ultra-faint galaxies. We fit this slope from the UV spectrum between rest-frame 
wavelength $0.15~\rm \mu m - 0.30~\rm \mu m$. At $z\sim5$ ultra-faint galaxies ($m = 30 - 33$)
have $\beta\approx -2.48\pm0.04$, while at $z\sim8$, $\beta\approx -2.62\pm0.04$.
The errors are the rms of the data, however, considering the dispersion of the escape fraction that is 
not modeled in simulations and the uncertainty on the star formation mode (constant star formation 
rate or instantaneous burst), the actual slope may span a larger range. 

As a caveat, results of Fig. \ref{halomass} and Fig. \ref{metallicity} for the faintest among the observable 
galaxies may be somewhat uncertain due to numerical resolution effects that are difficult to quantify 
at this stage. For example, \citet{2013MNRAS.429.2718S} pointed out that metallicity could be underestimated 
by 0.1 dex due to poor resolution.

\begin{figure}
\centering{
\includegraphics[scale=0.4]{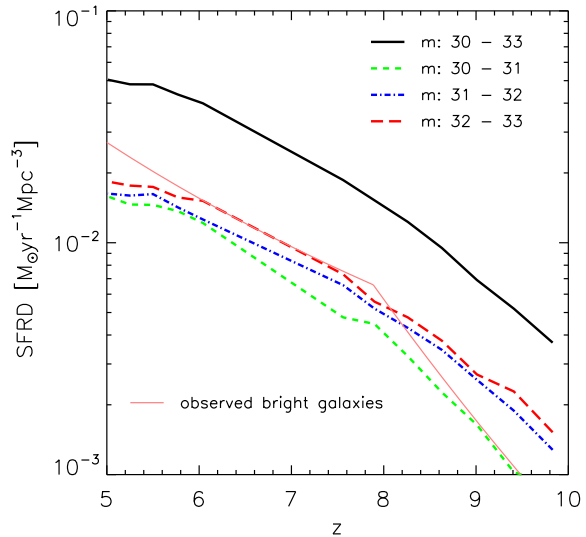}
\caption{Star formation rate density of ultra-faint galaxies in various magnitude bins.
}
\label{SFRD}
}
\end{figure}

We plot the star formation rate density (SFRD) of the ultra-faint galaxies in different magnitude bins as a function of 
redshift in Fig. \ref{SFRD}. The SFRD of galaxies with $M_{\rm AB} \simgt -15$ is calculated using the power-law 
extrapolation discussed in Sec. \ref{number} together with the SFR - $M_{\rm AB}$ relation obtained from simulations
which does not vary above and below this magnitude.
In the same panel, we also plot the SFRD that fits the observed bright ($m < -17.7$) galaxies for
$z \simlt 8$ ($\propto(1+z)^{-3.6}$) and $z \simgt 8$ ($\propto(1+z)^{-11.4}$)
by the thin line \citep{2013ApJ...773...75O}.
The ultra-faint galaxy SFRD is about $2-4$ times that for the currently known bright galaxies at any epoch. 
Thus, we are led to conclude that possibly the FFs project could unveil for the first time galaxies that are the major reionization sources.
It might also lend further support to the idea that the majority of star-forming galaxies are ultra-faint systems and are still below the 
detection limit of current surveys.

\begin{figure}
\centering{
\includegraphics[scale=0.4]{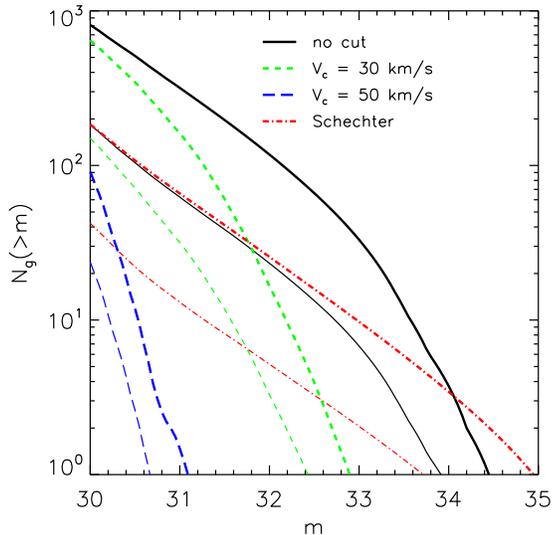}
\caption{Number of ultra-faint galaxies expected to be observed behind Abell 2744 (thin lines) and all six lenses (thick lines) 
in case different circular velocity cuts are adopted.
}
\label{feedback}
}
\end{figure}
\section{Radiative feedback effects }
In Sec. \ref{number} we have predicted the number of ultra-faint galaxies that would be detected
through gravitational lensing in the absence of radiative feedback.
However, if such faint galaxies are either not detected or their number is much smaller than expected, this would provide key missing information on the suppression of star formation in small galaxies by radiative feedback, i.e. photo-evaporation and/or heating of their gas by the cosmic UV background radiation

As a preliminary test, we assume halos below a certain circular velocity, $v_c$, do not form stars.
The results are plotted in Fig. \ref{feedback}. If halos with $v_c < 30(50)$~km s$^{-1}$ are starless, 
we should not observe galaxies fainter than $m\approx 33(31)$ in the FFs project. 

The LF slope, $\alpha$, contains observational uncertainty.  \citet{2011MNRAS.414..847S} simulations predict 
$\alpha \sim -2.0$ even for galaxies at $z\sim5$;
however, the slope derived from observations of bright galaxies
is shallower, $\alpha = -1.66\pm0.09$ at $z\sim5$ \citep{2007ApJ...670..928B}.
To check if our conclusions hold when adopting the Schechter LF that fits observations,
we plot 
the number of faint galaxies by using the
Schechter LF \citep{2014arXiv1403.4295B}. 
Even in this case we confirm that it is 
still possible to detect galaxies as faint as $m \approx 33-34$.

The shortage of small galaxies could also arise from modification of the power spectrum in WDM cosmologies \citep{2013MNRAS.435L..53P}. It will be possible to break this degeneracy by future observations that directly constrain the DM halo abundance. 

\section{Conclusions}
In this {\it Letter} we have predicted the properties of ultra-faint galaxies that should be detectable through the gravitational lensing in the FFs project. Using Bradac's FFs lens models, we estimated that in the most optimistic case galaxies as faint as $m \approx 33 - 34$ should be detected. Most ultra-faint galaxies (intrinsic $m=30 - 33$) are located in host halos with mass $\approx 10^{9}~M_\odot$, where the galaxy formation is believed to be sensitive to the feedback. Therefore the FFs project may provide an unique chance to directly assess the impact of radiative feedback in the early universe for the first time. 
We have also found that in $5 < z < 10$, the SFRD in galaxies between $m=30 - 33$ is about $2-4$ times the SFR of all high-$z$ bright galaxies observed so 
far. 
This implies that with FFs we are on the verge to unveil the major sources of cosmic reionization. Compared to bright galaxies, the abundance of ultra-faint galaxies is very sensitive to the effects of radiative feedback which suppresses the formation of stars in small halos: for example, if galaxy formation is suppressed in halos with circular velocity $v_c < 50$~km s$^{-1}$, we should not observe galaxies fainter than $m=31$, providing a stringent test to feedback models. 

\section*{Acknowledgements}
This work utilizes gravitational lensing models produced by PIs Bradac, Ebeling, Merten \& Zitrin, Sharon, and Williams funded as part of the HST Frontier Fields program conducted by STScI. STScI is operated by the Association of Universities for Research in Astronomy, Inc. under NASA contract NAS 5-26555. The lens models were obtained from the Mikulski Archive for Space Telescopes (MAST).


\end{document}